\documentclass[prb,showpacs,aps,twocolumn]{revtex4} 
\usepackage{amsmath}
\usepackage{epsfig}

\begin{document}
\title{Excitons in coupled InAs/InP self-assembled quantum wires}
\author{Y. Sidor}
\author{B. Partoens}
\author{F. M. Peeters}
\affiliation{Departement Fysica, Universiteit Antwerpen (CGB),
Groenenborgerlaan 171, B-2020 Antwerpen, Belgium}
\author{T. Ben}
\author{A. Ponce}
\author{D. L. Sales}
\author{S. I. Molina}
\affiliation{Departamento de Ciencia de los Materiales e I. M. y
Q. I., Universidad de C\'{a}diz, Puerto Real, C\'{a}diz, Spain}
\author{D. Fuster}
\author{L. Gonz\'{a}lez}
\author{Y. Gonz\'{a}lez}
\affiliation{Instituto de Microelectr\'{o}nica de Madrid
(CNM-CSIC), Isaac Newton 8, 28760 Tres Cantos, Madrid, Spain}
\date{\today}

\begin{abstract}

Optical transitions in coupled InAs/InP self-assembled quantum
wires are studied within the single-band effective mass
approximation including effects due to strain. Both vertically and
horizontally coupled quantum wires are investigated and the ground
state, excited states and the photoluminescence peak energies are
calculated. Where possible we compare with available
photoluminescence data from which it was possible to determine the
height of the quantum wires. An anti-crossing of the energy of
excited states is found for vertically coupled wires signaling a
change of symmetry of the exciton wavefunction. This crossing is
the signature of two different coupling regimes.

\end{abstract}


\maketitle

\section{Introduction}

Since the early 1970's, after the proposal of Esaki and
Tsu~\cite{Esaki}, multiple coupled quantum well structures have
been intensively investigated theoretically and
experimentally~\cite{boek1}. Early theoretical works on coupled
quantum wells were used in the study of
conduction~\cite{conduction,Chuang} and phonon tunnelling in
superlattices~\cite{supperlattices}.

Technological progress in the growth technique of low-dimensional
nacrostructures has shifted the research to the optical properties
of coupled quantum wires (CQWRs). For example, Kern \emph{et.
al.}\cite{Kern} experimentally and theoretically proved that the
coupling of the confined plasmons in Ga$_{x}$In$_{1-x}$As quantum
wires is much more enhanced with respect to the one in coupled
quantum wells. Later, Weman \emph{et. al.}\cite{Weman} reported a
strong exciton binding energy enhancement in very narrow
GaAs/Al$_{x}$Ga$_{1-x}$As cylindrical quantum wire arrays as
compared to quantum wells. The binding energy of the exciton was
determined from the experimental excitonic transitions and
compared to calculated values. In another experimental and
theoretical work done by Weman \emph{et. al.}\cite{WemanKapon} the
electron coupling and tunneling in double V-grooved
GaAs/Al$_{x}$Ga$_{1-x}$As QWRs was reported. Recently, the
interest has shifted towards the study of the electrical and the
optical properties in coupled (and also stacked)
\emph{self-assembled} quantum wires (see, for example,
Refs.~\onlinecite{APLCSQWR1,APLCSQWR2,Kapon}).

A further reduction of the dimensionality of nanostructures was
realized through the growth of quantum dots. Coupled
self-assembled quantum dots have been the subject of several
scientific works\cite{Pryor,Szafran,Pertoff,Karen,Bayer,China}.
Pryor\cite{Pryor} calculated within the eight-band
\textbf{k}$\cdot$\textbf{p} model the transition between a quantum
wire and vertically coupled quantum dots.

In the present work we investigate InAs/InP quantum wires which
are promising candidates for aplications in telecommunications,
because they emit at the wavelengths 1.3 and 1.55 $\mu$m. It was
demonstrated\cite{ExpAPL1} that the growth conditions of the InP
buffer layer controls the surface rearrangement of the strained
InAs layer that is grown on top. Therefore, it is possible to
obtain either quantum dot or quantum wire structures for identical
InAs coverage and growth conditions. Recently, optical properties
of InAs/InP self-assembled quantum wires were studied
experimentally\cite{ExpAPL2,Exp}. The wires are oriented along the
[1$\bar{1}$0] direction and periodically arranged along the [110]
direction with period 18 nm\cite{Exp}. Their photoluminescence
(PL) spectrum consists of four peaks that is believed to
correspond to different heights of the wires.

The main goal of the present work is the investigation of exciton
coupling in self-assembled InAs/InP quantum wires studied
experimentally in Ref.~\onlinecite{Exp}. We consider rectangular
self-assembled InAs/InP CQWRs. Since self-assembled InAs/InP CQWRs
are formed by the Stranski-Krastanow growth mode, effects due to
strain must be included~\cite{my1}. The calculations are performed
within the single-band effective mass approximation and based on a
two-dimensional (2D) finite element technique, where the mass
mismatch between the barrier and the wire is included. We examine
the dependence of the single particle energies and the behavior of
the electron and hole density on the distance between two
vertically and two horizontally coupled quantum wires, the Coulomb
interaction energy and the exciton energy on the small distance
between two vertically coupled quantum wires. We also calculate
the PL transition energies in these structures as a function of
the height of the wires and compare them with available
experimental data. From this comparison we are able to deduce the
height of the coupled wires.

\section{Model Hamiltonian}

The experimentally grown InAs/InP quantum wires are oriented along
the [1$\bar{1}$0] direction and periodically arranged along the
[110] direction, with period 180 \AA~\cite{Exp}. The average
height and the width of the wires were determined from XTEM
measurements. We consider two 2D rectangular quantum boxes at a
distance $d$ from each other, each with height $h$ and width $w$,
as illustrated in Fig. 1. The rectangular shape of the wires
approximates the experimentally measured shape. For self-assembled
InAs/InP quantum wires we have typically $w$ $\gg$ $h$. Therefore
we consider the situation of two horizontal coupled wires placed
along the wire width direction ([110] direction) as corresponding
to the experimental situation\cite{Exp} (see Fig. 1(a)), and two
vertically coupled wires directed along the height direction (see
Fig. 1(b)).

The full Hamiltonian for an exciton in such CQWRs consists of the
single electron part $H_{e}$, the single hole part $H_{h}$ and the
Coulomb interaction term between the electron-hole pair

\begin{equation}
H=H_{e}+H_{h}-\frac{e^{2}}{\varepsilon |\textbf{r}|},
\end{equation}
where \emph{e} is the free-electron charge, $\textbf{r}$ =
$\textbf{r}_{\textbf{e}}-\textbf{r}_{\textbf{h}}$ denotes the
relative distance between the electron (e) and the hole (h) and
$\varepsilon$ is the dielectric constant taken as the average
value of the wire and the barrier~\cite{my1}. According to the
effective-mass theory  the Schr\"{o}dinger equation for the
exciton can be written as

\begin{equation}
H\Psi(\textbf{r}_{\textbf{e}},\textbf{r}_{\textbf{h}})=E\Psi(\textbf{r}_{\textbf{e}},\textbf{r}_{\textbf{h}}).
\end{equation}

We assume that the conduction band and the valence band are
decoupled, which is a reasonable approximation for the considered
wires. The minimum of the conduction band and the maximum of the
valence band are localized around the $\Gamma$-point of the wire.
Along the wire growth direction (here taken to be the
$z$-direction) there is no confinement for the particles in the
wires, while in the $xy$-plane the electron and hole are strongly
confined. For this reason we are allowed to separate the
$z$-motion from the $xy$-motion. Further, we assume the Coulomb
interaction term as a perturbation, so that we can separate the
electron and hole wavefunctions. Next, we introduce the relative
coordinates $z$ = $z_{e}$-$z_{h}$ and the center-of mass (CM)
coordinates Z = ($m_{e}$$z_{e}$+$m_{h}$$z_{h}$)/$M$ which allows
us to write the solution to Eq. (2) as
\begin{eqnarray}
\Psi(\textbf{r}_{\textbf{e}},\textbf{r}_{\textbf{h}})=\Psi_{e}(x_{e},y_{e})\Psi_{h}(x_{h},y_{h})
\varphi(z)\exp(iK_{\mathrm{CM}}Z) ,
\end{eqnarray}
where $K_{\mathrm{CM}}$ is the momentum of the center-of mass and
$M$ = $m_{e}$+$m_{h}$ is the total mass. To calculate the electron
and hole energies, densities and the Coulomb interaction between
the particles we first solve, within the single band effective
mass approximation, the single-particle 2D Schr\"{o}dinger
equation in the $xy$-plane

\begin{subequations}
\begin{equation}
H_{e}\Psi_{e}(x_{e},y_{e})=E_{e}\Psi_{e}(x_{e},y_{e}),
\end{equation}
\begin{equation}
H_{h}\Psi_{h}(x_{h},y_{h})=E_{h}\Psi_{h}(x_{h},y_{h}),
\end{equation}
\end{subequations}
and subsequently an effective one-dimensional equation for the
motion in the $z$-direction where the Coulomb interaction is
included perturbatively

\begin{eqnarray}
\left[-\frac{\hbar^{2}}{2\mu_{z}}\nabla_{z}^{2}+
\frac{\hbar^{2}K_{CM}^{2}}{2M}+U_{eff}(z)\right]\varphi(z)
 = E_{C}\varphi(z), 
\end{eqnarray}
where $\mu_{z}$ denotes the reduced mass of the exciton along the
wire axis, $U_{eff}(z)$ is the effective potential and $E_{C}$ is
the Coulomb energy.

\begin{figure}
\vspace{0.0cm} \ \epsfig{file=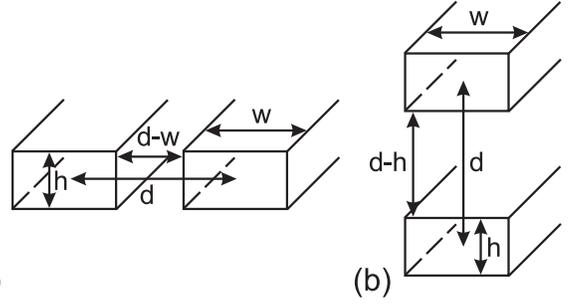,width=8 cm} \vspace{-0cm} \
\caption{Theoretical model of horizontally coupled [(a)] and
vertically coupled [(b)] rectangular quantum wires with height
$h$, width $w$, and distance $d$ between the wires.}
\end{figure}

We consider both heavy-hole (hh) and light-hole (lh) states.
Different effective masses of the particles are assumed inside and
outside the CQWRS. The single-particle 2D Hamiltonian for the
electron, and the two different holes (h) in the presence of
strain are given by
\begin{subequations}
\begin{eqnarray}
H_{e}&=&-\nabla_{xe}\frac{\hbar^{2}}{2m^{\ast}_{e}(x,y)}\nabla_{xe}
-\nabla_{ye}\frac{\hbar^{2}}{2m^{\ast}_{e}(x,y)}\nabla_{ye}\nonumber\\
&&+E_{ce}(x,y)+a_{c}\varepsilon_{hyd}(x,y),
\end{eqnarray}
\begin{eqnarray}
H_{h}&=&-\nabla_{x}\frac{\hbar^{2}}{2m^{\ast}_{h}(x,y)}\nabla_{xh}\nonumber\\
&&-\nabla_{yh}\frac{\hbar^{2}}{2m^{\ast}_{h}(x,y)}\nabla_{yh}+V_{vh}(x,y),
\end{eqnarray}
\end{subequations}
where $m^{\ast}_{e}(x,y)$ is the effective mass of the electron,
$E_{ce}(x,y)$ is the unstrained conduction band offset, $a_{c}$ is
the hydrostatic deformation potential for the conduction band,
$\varepsilon_{hyd}(x,y)$ is the hydrostatic strain,
$m^{\ast}_{h}(x,y)$ is the effective mass of the hole,
$V_{vh}(x,y)$ denotes the confinement potential of the hole due to
the band offsets and strain. The heavy-hole and light-hole
confinement potentials are obtained from the Pikus-Bir strain
Hamiltonian by its value in the center of the wire (see for more
details Ref.~\onlinecite{my1}). The effect of
heavy-hole-light-hole mixing is assumed to be negligible. From our
previous study on InAs/InP quantum wire \cite{my1} we know that it
leads to an insignificant correction to the exciton energy even in
the presence of a magnetic field.

\begin{table}
\caption{\label{tab:table1} Material parameters for InAs/InP CQWRs
used in the calculations: lattice constant $a_{0}$, band gap
$E_{g}$, electron mass $m_{e}$, Luttinger parameters $\gamma_{1}$
and $\gamma_{2}$ (Ref.~\onlinecite{Parameter}), the hydrostatic
deformation potential for the conduction band $a_{c}$, the
deformation potentials of the valence band $a_{v}$, $b$ and $d$,
strain coefficients $C_{11}$ and $C_{12}$ and dielectric constant
$\varepsilon$.}
\begin{ruledtabular}
\begin{tabular}{cccccc}
Parameter&InAs&InP \\
\hline
$a_{0}$(\AA) & 6.058 & 5.869 \\
$E_{g}$(eV) & 0.417 & 1.424 \\
$m_{e}$(m$_{0}$) & 0.023 & 0.077 \\
$\gamma_{1}$ & 20 & 5.08 \\
$\gamma_{2}$ & 8.5 & 1.6 \\
$a_{c}$(eV) & -5.08 & -- \\
$a_{\upsilon}$(eV) & 1 & -- \\
$b$(eV) & -1.8 & -- \\
$d$(eV) & -3.6 & -- \\
$C_{11}$(GPa) & 83.29 & -- \\
$C_{12}$(GPa) & 45.26 & -- \\
$\varepsilon$ & 15.15 & 12.5 \\
\end{tabular}
\end{ruledtabular}
\end{table}

The input parameters (see Table I) used for our simulations are
taken from Ref. \onlinecite{my1}, except for the mass of the
heavy-hole and the light-hole, which is taken the same in each
direction~\cite{my2}

\begin{subequations}
\begin{eqnarray}
\frac{m_{0}}{m^{\ast}_{ hh}} = \gamma_{1}-2\gamma_{2},
\end{eqnarray}
\begin{eqnarray}
\frac{}{}\frac{}{}\frac{}{} \frac{m_{0}}{m^{\ast}_{ lh}} =
\gamma_{1}+2\gamma_{2},
\end{eqnarray}
\end{subequations}
where $\gamma_{1}$ and $\gamma_{2}$ are Luttinger parameters (see
Ref.~\onlinecite{Parameter}), and $m_{0}$ is the vacuum electron
mass. The numerical calculations are based on the finite element
technique on a variable size grid.

\section{Comparison with experiment}\

\begin{figure}
\vspace{-0.5cm} \ \epsfig{file=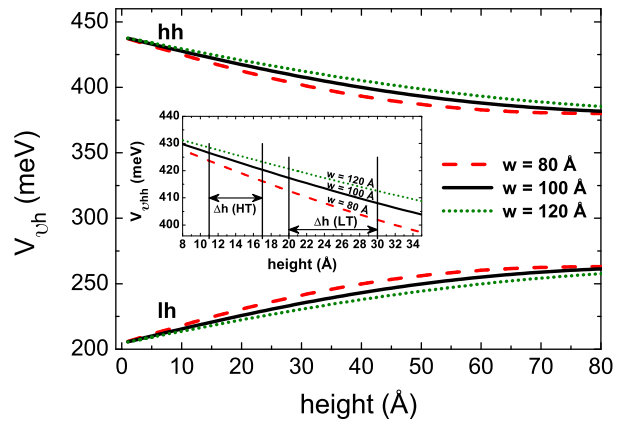,width=9.2 cm} \vspace{-0.8cm}
\ \caption{(color online) The top of the holes (h) potentials
calculated from the strain Hamiltonian at $x$ = 0 and $y$ = 0 as
function of the height of InAs/InP self-assembled CQWRs. The
dashed (red), full (black), and dotted (green) curves correspond
to the heavy-hole (hh) and the light-hole (lh) potentials of the
CQWRs with width $w$ = 80 \AA, $w$ = 100 \AA$ $ and $w$ = 120 \AA,
respectively. The inset shows the heavy-hole band offsets for the
High T (HT) and Low T (LT) samples (see  Ref. \onlinecite{Exp}).}
\end{figure}

In Ref. \onlinecite{Exp} the photoluminescence carrier
recombination in InAs/InP self-assembled wires was reported. The
PL spectra of two adjacent emission peaks of the High T (HT) and
Low T (LT) samples were investigated. The HT and LT samples were
grown at different substrate temperature and correspond to CQWRs
with different heights. The observed energy distance between these
two peaks for both the HT and LT samples correspond to 1 ML (1 ML
= 3 \AA) height fluctuation. The average geometric values of the
CQWRs were measured by XTEM\cite{Exp}.

\begin{figure} [b]
\vspace{-0.5cm} \ \epsfig{file=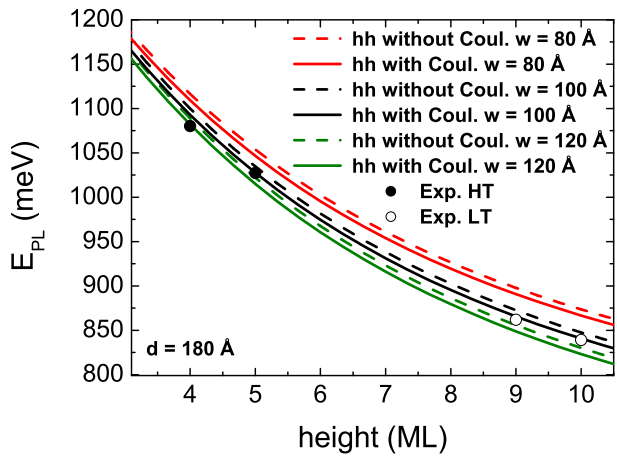,width=9.5cm} \vspace{-0.8cm}
\ \caption{(color online) PL peak energies as a function of the
InAs/InP CQWRs height. The full and dashed lines are the
theoretical calculations for the heavy-hole excitons with and
without taking into account the Coulomb interaction, respectively,
for the InAs/InP CQWRs with width $w$ = 80 \AA$ $ (red curves),
$w$ = 100 \AA$ $ (black curves) and $w$ = 120 \AA$ $ (green
curves). The full and open circles correspond to the experimental
data for HT and LT samples, respectively.}
\end{figure}

\begin{figure*} [t]
\epsfig{file=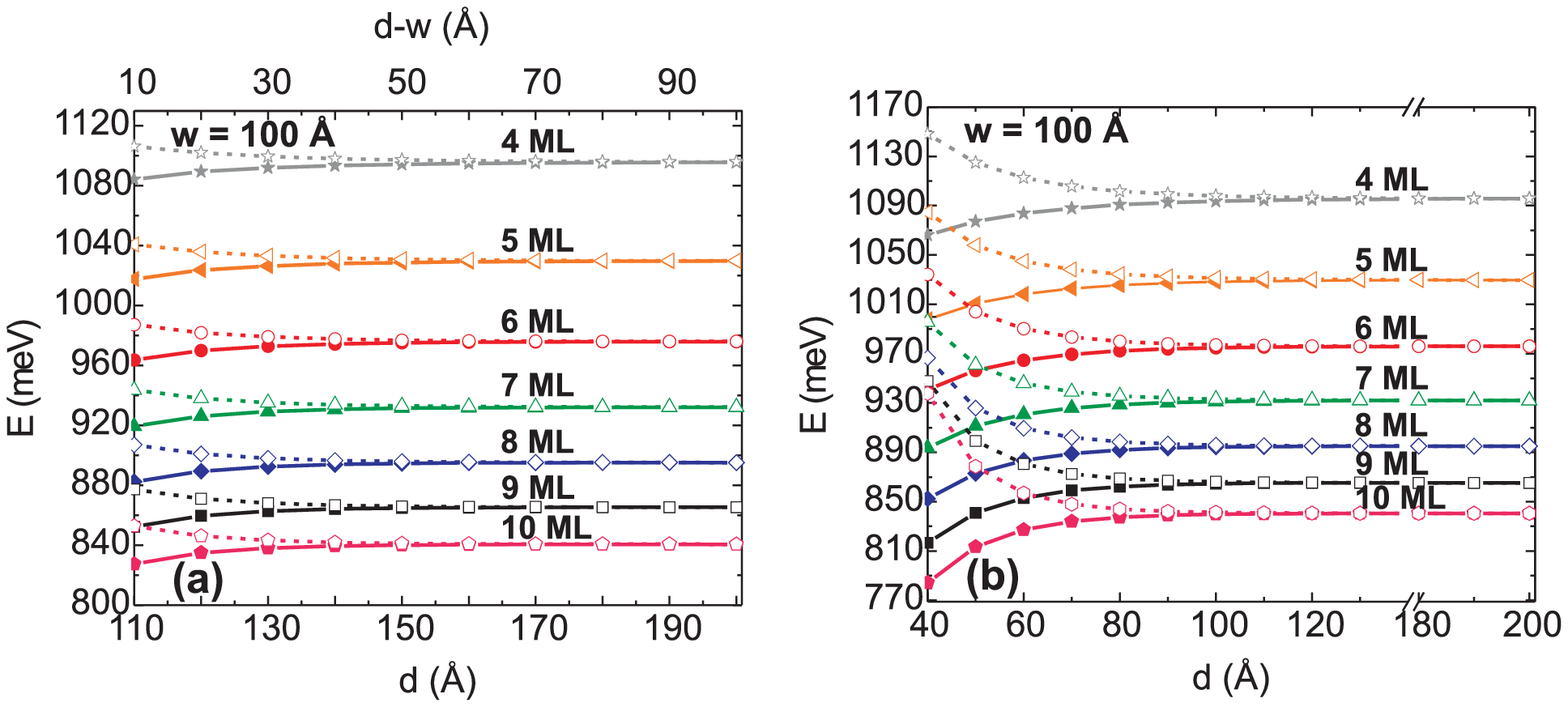,width=17.5 cm} \caption{(color online) The
heavy-hole exciton ground and the first excited state energies as
a function of distance $d$ in InAs/InP horizontally [(a)] and
vertically [(b)] CQWRs. The top scale, $d-w$, in the graph [(a)]
is the distance between the edges of two horizontally CQWRs (see
Fig. 1(a)). The full curves with the closed symbols and dashed
curves with the open symbols correspond to the ground and the
first excited state energies of the exciton, respectively. The
curves with the stars (grey), rotated triangles (orange), circles
(red), triangles (green), diamonds (blue), squares (black) and
pentagons (pink) correspond to the CQWRs with the height of 4 to
10 ML (1 ML is 3 \AA) in steps of 1 ML and width of 100 \AA,
respectively.}
\end{figure*}

In order to compare our calculations with the experimental data for
the HT and LT samples we consider the model of two horizontally
CQWRs, as mentioned in previous section. We used the heavy- and
light-hole band offsets depicted by the dashed, full, and dotted
curves in the inset of Fig. 2. The different curves correspond to
different widths of the CQWRs for both heavy- and light-holes.
Strain splits the heavy- and light-hole bands by a value which
depends on the matrix elements of the Pikus-Bir-Hamiltonian which
depend on the dimension of the quantum wire\cite{my1}. From Fig. 2
we notice that the heavy-hole curves are above the light-hole curves
in the whole range of heights up to 80 \AA$ $ and for three
different widths: $w$ = 80 \AA, $w$ = 100 \AA$ $ and $w$ = 120 \AA.
Therefore, we conclude that the heavy-hole state is the ground
state, as in the case of a strained quantum well. Further, in Fig.
3, we compare the calculated PL peak energies as a function of the
height of the InAs/InP CQWRs for three different widths with the
experimental PL energies of the HT and LT samples (indicated by the
symbols in Fig. 3). The distance between the quantum wires is fixed
to $d$ = 180 \AA. We are allowed to move the experimental points
along the height direction with a step of 1 ML to find the optimal
agreement with the theoretical curves. The results for the Coulomb
interaction included (when both electron and heavy-hole are in the
ground state) fit best the experimental points (see full curves in
Fig. 3), when the two peaks of the HT sample correspond to wire
heights of 4 and 5 ML, and the two peaks of the LT sample correspond
to heights of 9 and 10 ML. These height values of 4 and 5 ML, and
the values of 9 and 10 ML agree with those from XTEM measurements,
where the average value of the height for the HT was found to be $h$
= 14 $\pm$ 3 \AA,$ $ for the LT $h$ = 25 $\pm$ 5 \AA$ $, and the
average width was found to be $w$ = 100 $\pm$ 20 \AA.

\section{Inter-wire coupling: a theoretical investigation}

\begin{figure*}
\vspace{0cm} \ \epsfig{file=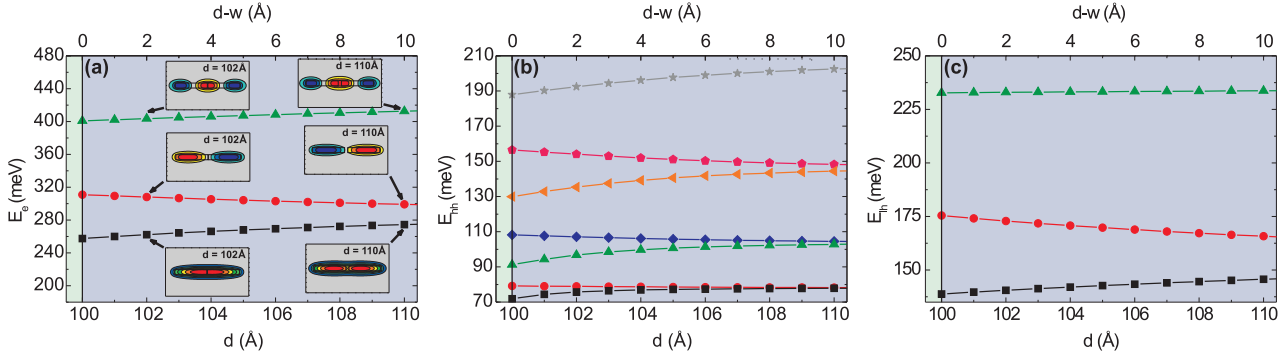,width=17.5 cm} \vspace{0cm} \
\caption{(color online) The energy for the electron [(a)] ground
state and the next two excited states, the heavy-hole [(b)] ground
state and the next six excited states and the light-hole [(c)]
ground state and the next two excited states as a function of
distance $d$ between the wires in InAs/InP horizontally CQWRs with
height $h$ = 27 \AA$ $ and width $w$ = 100 \AA. The top scale,
$d-w$, in the graphs [(a)], [(b)] and [(c)] is the distance
between the edges of two horizontally CQWRs (see Fig. 1(a)). The
insets in the graph [(a)] show the contour plots of the electron
wavefunctions for $d$ equal to 102 and 110 \AA.}
\end{figure*}

\begin{figure*}
\vspace{0cm} \ \epsfig{file=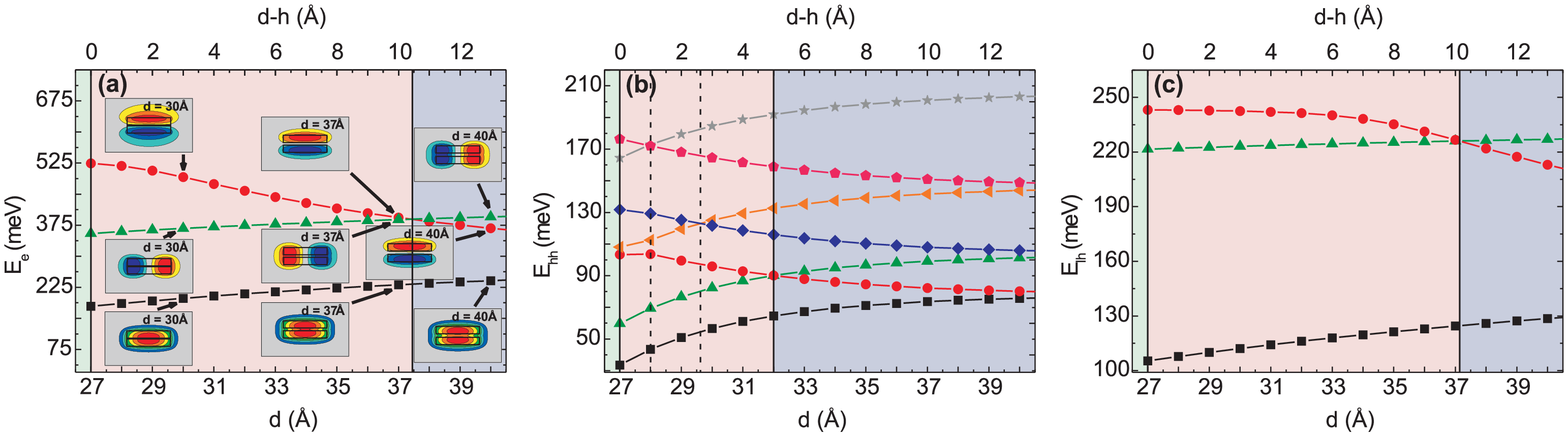,width=17.5 cm} \vspace{0cm} \
\caption{(color online) The same as Fig. 5 but now for vertically
CQWRs. The top scale, $d-h$, in the graphs [(a)], [(b)] and [(c)]
is the distance between the edges of two vertically CQWRs (see
Fig. 1(b)). The insets in the graph [(a)] show the contour plots
of the electron wavefunctions when $d$ is 40 \AA, 37 \AA$ $ and 30
\AA.}
\end{figure*}

Here we investigate the coupling for two wires when placed along
the wire width and the wire height direction, \emph{i.e.} the
horizontally and vertically CQWRs (see Fig. 1). In Fig. 4 we
examine the dependence of the energy of the ground and the first
excited state of the heavy-hole exciton as a function of the
distance between the wires. We study this dependence when the
height of the wires vary from 4 up to 10 ML (the width of the
wires is fixed to 100 \AA), which is an experimental range of
heights, as known from a previous PL study\cite{my3}. Fig. 4(a)
shows the results for the horizontally CQWRs where heavy-hole
exciton ground state energy slightly decreases as the
interdistance becomes smaller, while the first exciton excited
state energy slowly increases, as expected. For large
interdistances the energy levels become twofold degenerate. The
energy of the bound state (the ground state) decreases and the
energy of the anti-bounding state (the first excited state)
increases due to the lifting of the degeneracy of the two-fold
ground state. It is also clear that there is no evidence of
coupling in InAs/InP self-assembled wires when the interdistance
between the two wires is 180 \AA$ $ and therefore the results
presented in Fig. 3 are identical to those for a single
(\emph{i.e.} uncoupled) quantum wire. Coupling sets in when $d$
becomes roughly smaller than 160 \AA. Similar to the horizontally
CQWRs, in vertically CQWRs the heavy-hole exciton ground state
(first excited state) energies decrease (increase) as $d$
diminishes (see Fig. 4(b)). Here coupling is found when $d$ is
smaller than 120 \AA, and a stronger dependence at small values of
$d$ for the exciton ground and the first excited states is
noticed.

In Figs. 5 and 6 we examine the dependence of the energy for the
individual particles at smaller inter-wire distance $d$, up to the
moment when two wires in the coupled wire touch each other which
occurs for $d$ = 100 \AA$ $ in case of the horizontally CQWRs and
when $d$ = 27 \AA$ $ (we consider the height $h$ to be 9 ML = 27
\AA$ $ in Figs. 5 and 6) for the vertically CQWRs (see the left
vertical straight lines at $d-w$ and $d-h$ equal to 0 \AA $ $ in
Figs. 5 and 6, respectively). Let us consider first the single
particles for the horizontally CQWRs (see Figs. 5(a) - 5(c)). For
all particles (electron, heavy- and light-hole) we observe a
continuous energy splitting until $d$ reaches 100 \AA $ $ ($d-w$ =
0 \AA), and the wavefunctions simply merge into the wavefunction
of a single wire (compare the wavefunctions for the electron
ground and excited state energies in the inset of Fig. 5(a), when
$d$ is equal to 102 and 110 \AA). Besides, we can see that the
energy splitting between the bound state and the anti-bounding
state is larger for the electron and the light-hole (see Figs.
5(a) and 5(c)) as compared to the heavy-hole states (see Fig.
5(b)). The larger sensitivity of the electron and the light-hole
to the variation of the inter-wire distance is because of their
lighter effective mass which is about 10 times less in the wire
and at least 5 times less in the barrier than the heavy-hole
effective mass.

Interesting new effects arise when we consider vertically coupled
InAs/InP quantum wires. Again, the electron ground state
wavefunction merges together (compare the electron ground state
wavefunction in Fig. 6(a), when $d$ equals to 40, 37 and 30 \AA).
Meanwhile, the ground state energy of the electron decreases with
decreasing $d$, as we have previously shown in Fig. 4(b). For the
first excited state energy of the electron, when the distance
between the wires becomes smaller (see the curve (red) with the
circles in Fig 6(a)), the electron first excited state energy
increases, as it is shown in Fig. 4(b) up to a certain value of
$d$ between 37 and 38 \AA$ $ (see the right vertical straight line
in Fig. 6(a)). Starting from this value and up to $d$ equal to
$h$, it is energetically more favorable for the electron to jump
to another state, whose energy decreases (see the curve (green)
with the triangles in Fig. 6(a)). Also the symmetry of the
electron wavefunction of the first two excited states in the
transverse directions of the wires have a different behavior with
variation of the inter-wire distance. For this reason there is a
crossing between the excited states for the electron in the
vertically coupled wires, contrary to the horizontally coupled
wires where the electron wavefunction symmetry of the excited
states remains the same with the varying $d$ (see Fig. 5(a) and
Fig. 6(a)). Furthermore, in the inset of Fig. 6(a), where the
contour plots of the electron wavefunctions are shown, before and
after the crossing the symmetry with respect to the nodal line
changes for the first excited state and in the opposite way for
the second excited state. Note that when $d$ is between 27 and
roughly 38 \AA$ $, the wavefunction for the first two excited
states has the same character as for a rectangular single wire. A
similar crossing for the heavy-hole (see the right vertical
straight line in Fig 6(b) when $d$ equals 32 \AA$ $) and for the
light-hole (see the right vertical straight line in Fig 6(c) when
$d$ is roughly 37 \AA) between the first and second excited states
is observed. Besides, for the heavy-hole we found crossings
between the fourth and fifth (see the dashed vertical line between
29 and 30 \AA) as well as between the sixth and seventh (see the
dashed vertical line placed at $d$ = 28 \AA) heavy-hole excited
states. The electron and the light-hole have only three bound
states. Note that for electron and holes in InAs/InP vertically
CQWRs three regions can be distinguished:

(1) coupled region where the wavefunctions of the lowest levels
behave as for a single rectangular wire. But energetically they
are already in the coupled regime because the levels are all split
(see the middle region in Figs. 6(a) and 6(c));

(2) coupled region where the energies and the densities of the
lowest levels behave as for the coupled wires (see the right
region in Figs. 6(a) and 6(c));

(3) decoupled region, where all levels are twofold degenerate
(when $d$ is large).

\begin{figure}
\vspace{-0.4cm} \ \epsfig{file=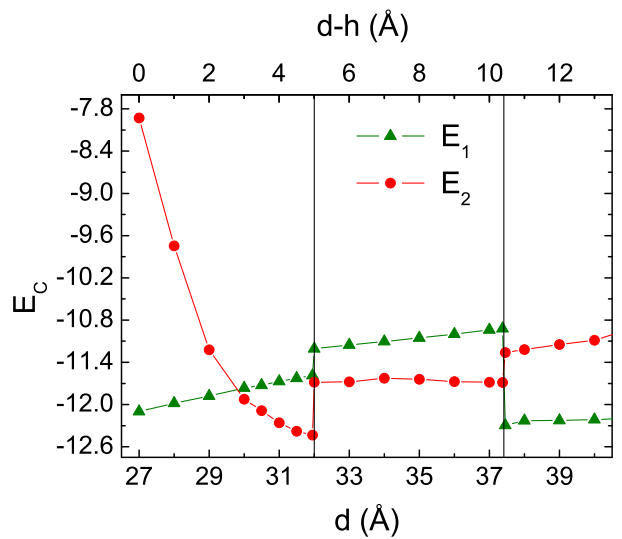,width=9.5cm} \vspace{-0.8cm}
\ \caption{(color online) The electron-heavy-hole Coulomb
interaction energy as a function of distance $d$ between the wires
in InAs/InP vertically CQWRs with height $h$ = 27 \AA$ $ and width
$w$ = 100 \AA. The curves with the triangles (green) and circles
(red) correspond to the Coulomb interaction energy between the
particles in the first and second excited state, respectively. The
top scale, $d-h$, in the graphs is the distance between the edges
of two vertically CQWRs (see Fig. 1(b)).}
\end{figure}

Next, we investigate the electron-hole Coulomb interaction energy
dependence on the distance $d$ between two vertically CQWRs. In
order to calculate the Coulomb energy of the first (second)
excited state we solve the 1D Schr\"{o}dinger equation for the
relative exciton motion in the $z$-direction (see Eq. (5)) with
the effective potential of the first (second) excited state
\begin{eqnarray}
U_{eff1(2)}(z)&=&\int\
dx_{e}dy_{e}dx_{h}dy_{h}|\Psi_{1(2)e}|^{2}|\Psi_{1(2)h}|^{2}\times\nonumber\\
&&\left(-\frac{e^{2}}{\varepsilon|\mathbf{r}_{e}-\mathbf{r}_{h}|}\right),
\end{eqnarray}
where $\Psi_{1(2)}$ is the electron and hole wavefunction of the
first (second) excited state. In Fig. 7 the Coulomb energies
between the electron and heavy-hole both in the first excited
state (see the curve with the triangles in Fig. 7) and in the
second excited state (see the curve with the circles in Fig. 7)
are shown. We found abrupt jumps for these two energies at the
crossing of the first two excited states for the electron at $d$ =
37.4 \AA$ $ and the heavy-hole $d$ = 32 \AA. Both energies
decrease continuously up to the crossing of the excited states of
the electron (see the right vertical line at 37.4 \AA$ $ in Fig.
7), meanwhile the absolute value of the Coulomb energy E$_{1}$ is
larger than the absolute value of the Coulomb energy E$_{2}$ of
the second excited state. After this crossing up to $d$ = 32 \AA$
$ the situation is opposite, \emph{i.e.}, $|$E$_{2}$$|$ $>$
$|$E$_{1}$$|$  even when the symmetry of the heavy-hole
wavefunctions of the first two excited states does not change (it
changes at $d$ = 32 \AA). This means a larger electron
contribution to the Coulomb interaction energy of the excited
states in comparison to the heavy-hole one. When the crossing
point of the heavy-hole excited states is reached (see the left
vertical line), the second abrupt jump in the Coulomb energy is
observed. However, because of the spill-over of the electron
wavefunction of the second excited state (there is no spill-over
effect observed for the heavy-hole excited states), the absolute
value of the Coulomb energy E$_{2}$ decreases after this crossing.
Therefore, the Coulomb energy E$_{2}$ becomes smaller in absolute
value than the value of E$_{1}$ for $d$ $<$ 30 \AA. Close and at
the crossing points, \emph{i.e.} $d$ $=$ 32 \AA$ $ and 37.4 \AA,
there will be a strong Coulomb interaction induced mixing of the
energy levels which should be included. For this reason we took
the electron (hole) wavefunction as a linear combination of the
first and the second excited state wavefunction
\begin{eqnarray}
\Psi_{e(h)} = (a\Psi_{1e(h)}+\Psi_{2e(h)})/\sqrt{1+|a|^{2}},
\end{eqnarray}
where 'a' is a weighting parameter which is taken as a variational
coefficient that minimizes the total exciton energy. In Fig. 8 we
plot the exciton energy dependence of the first two excited states
on the inter-wire distance $d$ between two vertically aligned
CQWRs. The results of the two different methods are presented: by
solving the single-particle Schr\"{o}dinger equations; and by
using the variational technique. In the first one, as was already
mentioned in Sect. II, the wavefunction in the $xy$-plane for the
first (second) excited state of the exciton is assumed as the
product of the wavefunction of the electron and the heavy-hole in
the first (second) excited state. Then, the total exciton energy
of the first (second) excited state equals $E_{Exc.,1(2)}$ =
$E_{e,1(2)}$+$E_{hh,1(2)}$+$E_{c,1(2)}$. In the second approach,
the total exciton wavefunction in the $xy$-direction is taken as
\begin{eqnarray}
\Psi(x_{e},y_{e},x_{h},y_{h},a,b) =
\Psi_{e}(x_{e},y_{e},a)\Psi_{h}(x_{h},y_{h},b),
\end{eqnarray}
using the new wavefunctions (see Eq. (9)). Then, we average the
kinetic part and the Coulomb interaction part in the exciton
Hamiltonian (see Eq. (1)). After this averaging we obtain the 1D
Schr\"{o}dinger equation for the relative exciton motion in the
$z$-direction, where the single particle energies $E_{e}$ +
$E_{hh}$ are now replaced by
$|a|^{2}$$E_{e,1}$+$E_{e,2}$+$|b|^{2}$$E_{hh,1}$+$E_{hh,2}$ and
the effective potential becomes now
\begin{eqnarray}
U_{eff}(z,a,b)&=&N\int
dx_{e}dy_{e}dx_{h}dy_{hh}(|a|^{2}|b|^{2}|\Psi_{1e}|^{2}|\Psi_{1h}|^{2}\nonumber\\
&&+|a|^{2}|\Psi_{1e}|^{2}|\Psi_{2h}|^{2}+|b|^{2}|\Psi_{2e}|^{2}|\Psi_{1h}|^{2}\nonumber\\
&&+|\Psi_{2e}|^{2}|\Psi_{2h}|^{2})\times\left(-\frac{e^{2}}{\varepsilon|\mathbf{r}_{e}-\mathbf{r}_{h}|}\right),
\end{eqnarray}
where $N = 1/((1+|a|^{2})(1+|b|^{2}))$. By numerically integrating
this equation and by minimizing the total energy we obtain the
exciton energy. From Fig. 8 we notice an anti-crossing for the
exciton excited state energies at $d$ = 37.4 \AA, where the
crossing  of the first two excited states for the electron was
found. However, at the point $d$ = 32 \AA$ $ both energies
slightly change their behavior. The slope of the curves of these
exciton energies change insignificantly (see the curves E$_{1}$
and E$_{2}$ near $d$ = 32 \AA$ $ in Fig. 8). Hence, the crossing
of the electron first and the second excited state energies leads
to the anti-crossing of the exciton first and the second excited
state energies, but the crossing of the heavy-hole first two
excited states does not affect much the exciton excited state
energies. Moreover, we see full agreement in the whole inter-wire
region $d$ between the energy calculated by the variational method
and between the first excited state energy of the exciton. It is
expected, since we always get the first excited state energy of
the exciton after minimization of the electron and heavy-hole
terms with the first and second excited state energies, and the
terms with the Coulomb interaction energy. However, near the
crossing of the excited states of the electron and the heavy-hole,
the exciton first excited state energy differs by less than 2 meV
from the one calculated from the variational technique (see the
insets (a) and (b) of Fig. 8). At the crossing points there are
abrupt jumps for the curve E$_{1}$, resulting from the jump in the
Coulomb interaction energy for the first excited state (see Fig.
7), while continuous behavior is found for the variational
calculation results (see the curve with the open squares in the
insets of Fig. 8). This is because near these crossings the
difference between the energies of the excited states of the
electron and the heavy-hole  becomes smaller than the difference
between the Coulomb interaction energies of the excited states.

\begin{figure}
\vspace{-0.4cm} \ \epsfig{file=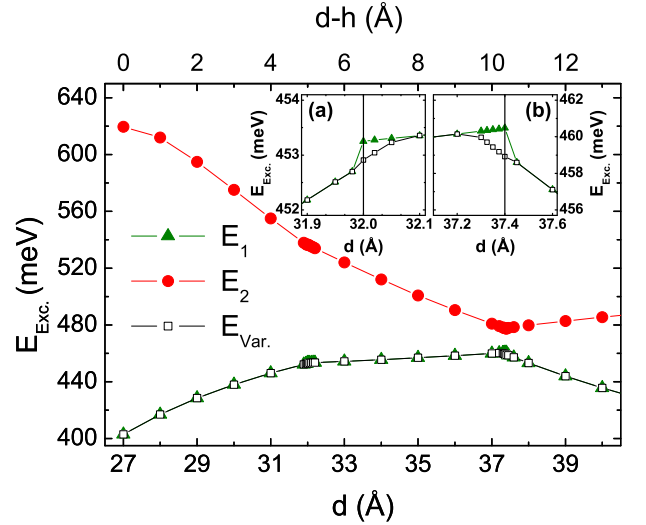,width=9.5cm} \vspace{-0.8cm}
\ \caption{(color online) The heavy-hole exciton energy as a
function of distance $d$ between the wires in InAs/InP vertically
CQWRs with height $h$ = 27 \AA$ $ and width $w$ = 100 \AA. The
curves with the full triangles (green) and full circles (red)
correspond to the exciton first and second excited state energies,
respectively. The curve with the open squares (black) correspond
to our variational calculation. The top scale, $d-h$, in the
graphs is the distance between the edges of the two vertically
CQWRs. The inset shows the exciton energy near the crossing of the
first and the second excited states for heavy-hole (a) and
electron (b)(see Figs. 6(a) and 6(b)).}
\end{figure}

\section{Conclusions}

We investigated the coupling in InAs/InP horizontally and
vertically CQWRs. The calculation are performed within the
single-band effective mass approximation. The strain effect on the
band offsets, the mass mismatch in the barrier and the wire, and
the Coulomb interaction between the electron-hole pair are
included. We found that the heavy-hole state is the ground state.

The PL energy for horizontally coupled InAs/InP wires was
calculated as function of the wire height at fixed width. From
comparison with the observed PL energies we derive the height of
the quantum wire which is in good agreement with those from the
experimental XTEM measurements. No inter-wire coupling of the
exciton in the experimentally grown InAs/InP self-assembled
quantum wires with period $d$ = 180 \AA$ $ was found. Such
coupling effects is predicted to show up when the distance $d$
between the wires is smaller than 160 \AA.

Numerical results for InAs/InP horizontally coupled system show
one coupling regime for all inter-wire distances between the
wires. However, for vertically CQWRs a crossing between the
excited states for the particles is predicted when the distance
between the wires approaches the value of the wires heights. Due
to this crossing two coupling regimes for InAs/InP vertically
CQWRs are found. In the first coupling regime the electron, heavy-
and light-hole densities for the lowest levels have the same
behavior as in a single square wire, while in the second coupling
regime both the energies and densities for the lowest levels act
as in ordinary coupled wires.

Anti-crossing for exciton excited state energies in vertically
CQWRs is predicted for the inter-wire distance where the crossing
of the first and the second excited states for the electron is
found. Exciton excited state energies are slightly affected when
the symmetry of the wavefunction of heavy-hole excited states in
vertically CQWRs changes. In a PL experiment the heavy-hole
exciton line is bright while the corresponding one for the
light-hole is dark due to a selection rule. Therefore, we did not
present results for the light-hole exciton.

\section{Acknowledgments}

This work was supported by the European Commission network of
excellence: SANDiE and the Flemish Science Foundation (FWO-Vl).
The authors thank M. Hayne and V. V. Moshchalkov for stimulating
discussions.

\end{document}